\documentclass[aps,prb,reprint,superscriptaddress]{revtex4-2}

\usepackage{bbm}
\usepackage{amssymb, amsmath,leftidx}
\usepackage{amsmath}
\usepackage{graphicx}
\usepackage{times}
\usepackage{CJK}
\usepackage{color}
\usepackage[normalem]{ulem}

\begin{document}


\begin{CJK*}{UTF8}{}
\CJKfamily{bsmi}

\title{Finite-size scaling analysis of two-dimensional deformed Affleck-Kennedy-Lieb-Tasaki states}



\author{Ching-Yu Huang (黃靜瑜)}
\affiliation{ National Center for Theoretical Sciences, Hsinchu 30013, Taiwan}
\affiliation{ Department of Applied Physics, Tunghai University, Taichung 40704, Taiwan}

\author{Yuan-Chun Lu (呂元鈞)}
\affiliation{ Department of Physics, National Tsing Hua University, Hsinchu 30013, Taiwan}

\author{Pochung Chen (陳柏中)}
\email{pcchen@phys.nthu.edu.tw}
\affiliation{ Department of Physics, National Tsing Hua University, Hsinchu 30013, Taiwan}


\date{\today}

\begin{abstract}
Using tensor network methods, we perform finite-size scaling analysis to study the parameter-induced phase transitions
of two-dimensional deformed Affleck-Kennedy-Lieb-Tasaki states.
We use higher-order tensor renormalization group method to evaluate the moments and the correlations.
Then, the critical point and critical exponents are determined simultaneously by collapsing the data.
Alternatively, the crossing points of the dimensionless ratios are used to determine the critical point,
and the scaling at the critical point is used to determine the critical exponents.
For the transition between the disordered AKLT phase and the ferromagnetic ordered phase,
we demonstrate that both the critical point and the exponents can be determined accurately.
Furthermore, the values of the exponents confirm that the AKLT-FM transition belongs to the 2D Ising universality class. 
We also investigate the Berezinskii-Kosterlitz-Thouless transition from the AKLT phase to the critical XY phase.
In this case we show that the critical point can be located by the crossing point of the correlation ratio.
\end{abstract}


\maketitle
\end{CJK*}

\section{Introduction}

In recent years, coarse-graining renormalization group methods for the tensor network have become 
the essential numerical tools to study classical and quantum lattice models.
One advantage is the ability to study systems in the thermodynamic limit.
However, it can be tricky to extract the critical properties of an infinite system,
due to the crossover to a mean-field like behavior when the system approaches the critical point \cite{Liu:2010dl}.
There are more complex tensor coarse-graining methods \cite{Evenbly:2015csa, Evenbly:2015ey, Yang:2017hj},
which can be used to extract conformal data such as the central charge and the scaling dimensions. 
But these methods are typically computationally more expensive. 
Another standard route to extract the critical properties is to calculate the physical properties of a finite-size system, 
then perform  finite-size scaling (FSS) analysis.
To the best of our knowledge, however, tensor network methods have rarely been used to perform FSS analysis for two-dimensional (2D) classical systems.
A new exceptions include the corner transfer matrix renormalization group study for 2D classical lattice models \cite{Nishino:1997kn},
and the tensor network based FSS analysis of the Fisher zeros \cite{Denbleyker:2014cda, Hong:2019cm}.
This is partly due to the lack of an efficient and accurate method to calculate the higher-order moments via tensor network methods,
while higher-order moments and their ratios are key ingredients of the FSS analysis.
Recently, an algorithm to calculate higher order moments via HOTRG is proposed in Ref.~\cite{Morita:2018gy}.
This motivates us to revisit the route of tensor network based FSS analysis.

In this work we demonstrate that the FSS analysis can be performed within the tensor network framework in a fashion that is similar to the Monte Carlo simulations.
In particular, we study the phase transitions  and the critical properties of the ``deformed-AKLT'' family of states on 2D square lattice.
One-dimensional (1D) Affleck, Kennedy, Lieb, and Tasaki (AKLT) state \cite{Affleck:1987jya, Affleck:1988hl} 
is originally constructed to understand Haldane's conjecture \cite{Haldane:1983ip} for the integer-spin chains.
Later, it becomes a canonical example for the concept of the symmetry protected topological order \cite{Pollmann:2010ih}.
It is straightforward to generalize the valence bond construction of the 1D AKLT state to higher dimensions.
In recent years, the 2D generalizations of the AKLT states and their parent Hamiltonians have been actively investigated 
to understand the nature of the symmetry protected topological order \cite{Chen:2011gt, You:2014fo, Wierschem:2016kv, Huang:2016bf, Pomata:2018ep}.
Specifically, the ``deformed-AKLT'' family of states 
is a two-parameter family of states on the 2D square lattice, which can be obtained by deforming locally the 2D AKLT state.
As one varies the parameters within the parameter space, the state exhibits parameter-induced phase transitions.
It is known that the phase diagram consists of a disordered AKLT phase, a ordered Ferromagnetic (FM) phase (or equivalently a N\'eel phase), 
and a critical XY phase \cite{Niggemann:2000ta, Pomata:2018ep}.

Specifically, we use higher order tensor renormalization group (HOTRG) \cite{Xie:2012iy} 
method to coarse-grain the tensor network and to calculate the correlations as well as the higher moments~\cite{Morita:2018gy}.
Then we perform FSS analysis based on the moments, the correlations, and their dimensionless ratios.
For the AKLT-FM transition, we are able to determine the critical point and the critical exponents accurately.
Furthermore, the values of the critical exponents confirm that the transition belongs to the 2D Ising universality class.
For the AKLT-XY transition, we show that the critical point can be located by the FSS analysis of the correlation ratio.

The manuscript is organized as follows:
In Sec.\ref{model_method} we briefly describe how to construct higher dimensional generalization of the AKLT state 
and how to construct the deformed-AKLT family of states on 2D square lattice.
We present our FSS analysis and the results for the AKLT-FM transition in Sec.\ref{AKLT-FM}.
Then in Sec.\ref{AKLT-XY} we investigate the BKT transition into the critical XY phase.
Finally in Sec.\ref{discussion} we discuss various aspects of the approaches used in this work.

\section{Model and Method\label{model_method}}

Our starting point is a generalized AKLT state on an arbitrary lattice, which admits a tensor network state representation.
Consider a lattice with  coordination number $q$. First we put a virtual spin-$\frac{1}{2}$ state at the end of each link. 
Then we project the $q$ virtual spin states on each vertex onto the subspace of spin-$\frac{q}{2}$ with the projector:
\begin{equation}
  \mathcal{P}_q= \sum c_{s} \left|\frac{q}{2}, s \right\rangle \langle s_1,s_2,...,s_q|,
\end{equation} 
where $s_i = \pm \frac{1}{2}$ and $s=\sum_i s_i \in [-\frac{q}{2}, \dots, \frac{q}{2}]$ are the virtual and physical spin index in their $S^z$ basis respectively,
and $c_{s}$ are the Clebsch-Gordan coefficients. Next we put a bond state $|\omega\rangle$ on each link $l$ of the lattice,
where the bond states $ | \omega \rangle$ is one of the Bell states
\begin{align}
& | \phi^{+} \rangle = | +\frac{1}{2}, +\frac{1}{2}\rangle + |-\frac{1}{2}, -\frac{1}{2}\rangle \notag \\
& | \phi^{-} \rangle = | +\frac{1}{2}, +\frac{1}{2}\rangle - |-\frac{1}{2}, -\frac{1}{2}\rangle  = I \otimes  \sigma^z | \phi^{+} \rangle    \notag \\
& | \psi^{+} \rangle = | +\frac{1}{2}, -\frac{1}{2}\rangle + |-\frac{1}{2}, +\frac{1}{2}\rangle = I \otimes  \sigma^x | \phi^{+} \rangle  \notag \\
& | \psi^{-} \rangle = | +\frac{1}{2}, -\frac{1}{2} \rangle - |-\frac{1}{2}, +\frac{1}{2}\rangle  =I \otimes  i\sigma^y | \phi^{+} \rangle.
\end{align} 
Here $\sigma^x$, $\sigma^y$, and $\sigma^z$ are Pauli matrices. The spin-$\frac{q}{2}$ AKLT state is then defined as
\begin{equation}
  \left|\Psi_{\text{AKLT}} \left(\frac{q}{2} \right) \right\rangle = \bigotimes_{v \in V}  ( \mathcal{P}_q)_v \bigotimes_{l \in L}  | \omega \rangle_l.
\end{equation} 
To construct the ``deformed-AKLT'' family of states, we apply the following diagonal, spin-flip invariant deformation
\begin{equation}
  D(\vec{a}) = \sum_{s= -q/2} ^{q/2} \frac{a_{|s|} }{c_s} | s \rangle \langle s|,
\end{equation} 
where $s$ is the physical spin index. The resulting family of states can hence be expressed as
\begin{equation}
  \left|\psi_{\text{AKLT}} \left(\frac{q}{2}, \vec{a} \right) \right\rangle =  \bigotimes_{v \in V}   \left( D(\vec{a}) \mathcal{P}_q \right)_v
   \bigotimes_{l \in L}  |\omega\rangle_l.
\end{equation}

\begin{figure}[t]
  \includegraphics[width=1.0\columnwidth]{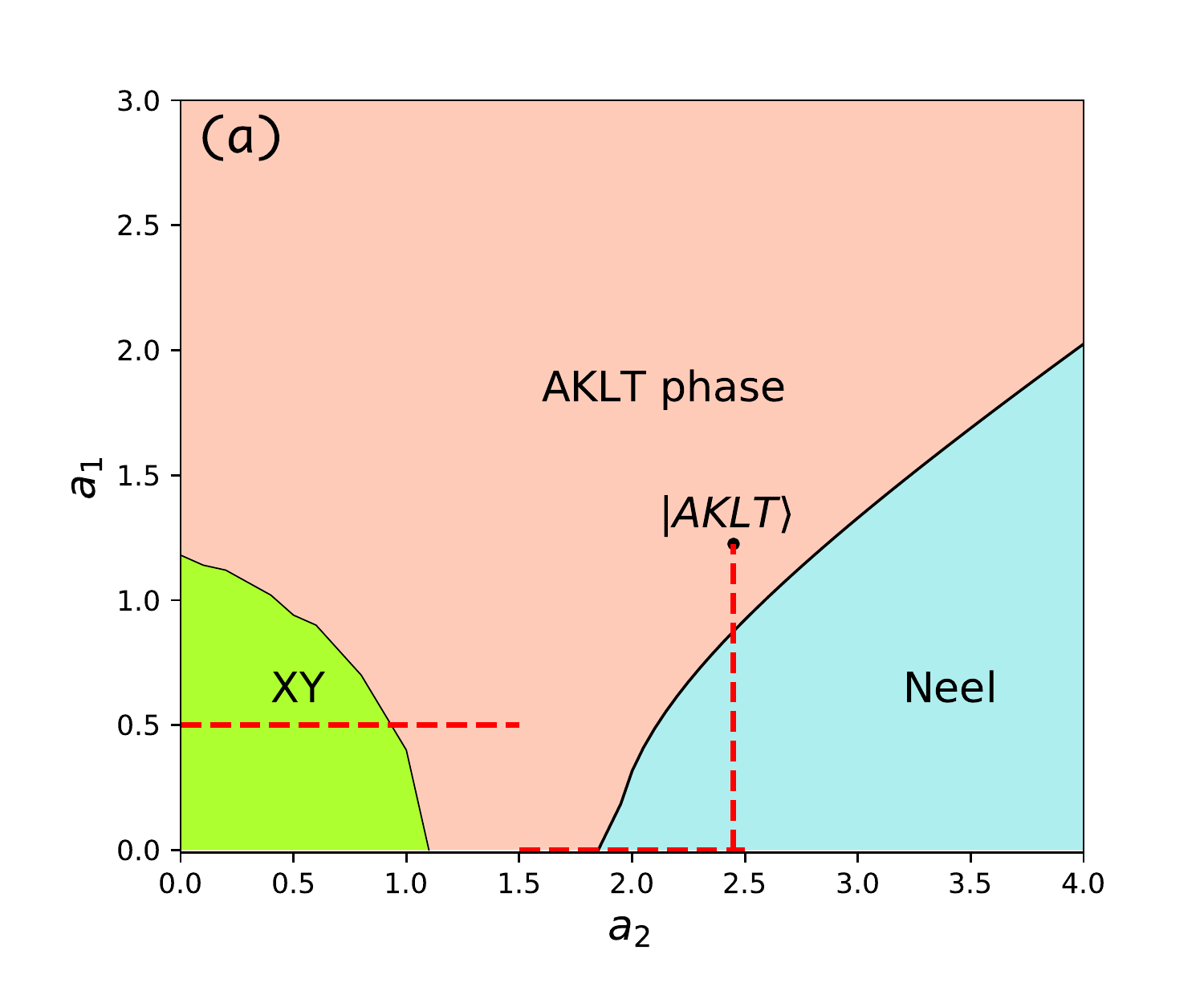}  
  \includegraphics[width=1.0\columnwidth]{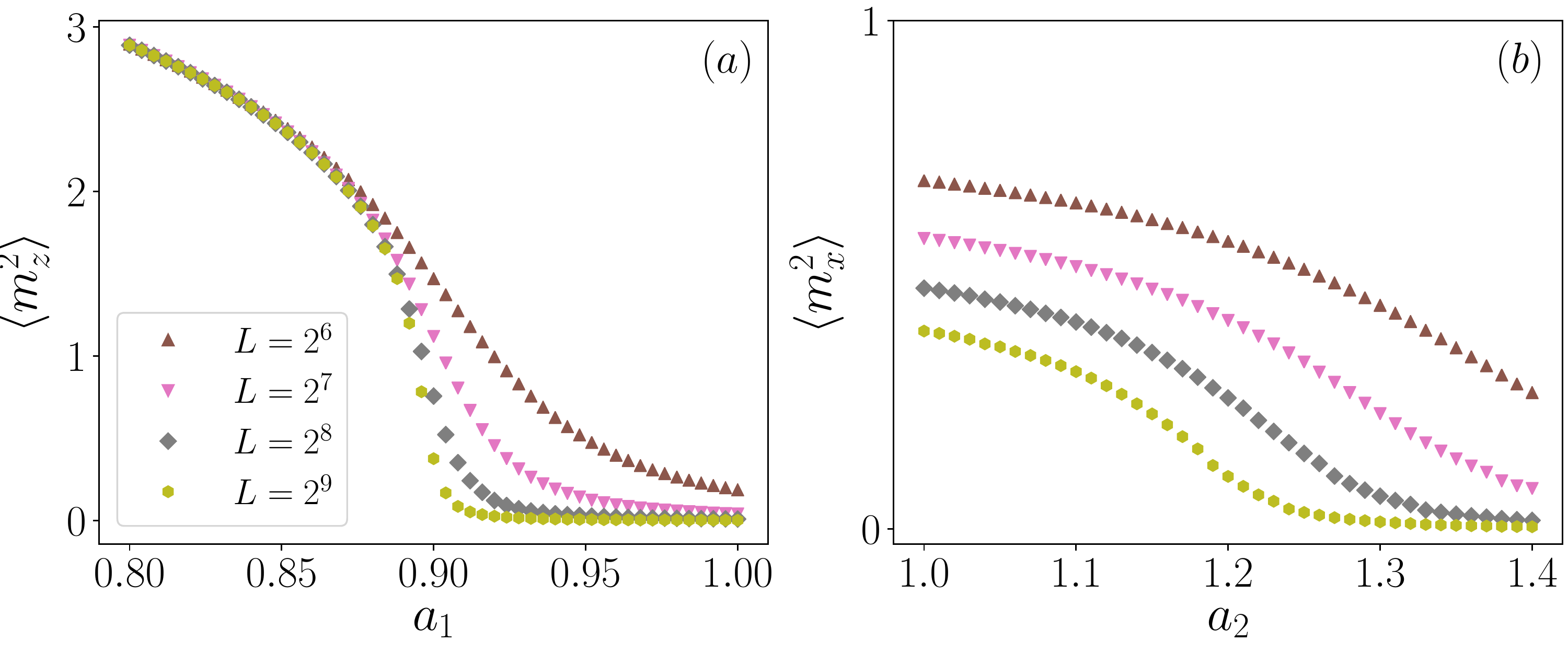}  
  \caption{(Color online) (a) Sketch of the phase diagram. Red dashed lines correspond to the three scans of the parameters performed in this work.
  (a) Second moment $\langle m^2_z \rangle$ as a function of  $a_1$ with $a_2=\sqrt{6}$ near the AKLT-FM transition. 
  (b) Second moment $\langle m^2_x \rangle$ as a function of $a_2$ with $a_1=0.5$ near the AKLT-XY transition.
  The linear system sizes are $L=64, 128, 256, 512$.}
  \label{fig:phase}
\end{figure}

We note in passing that on a bipartite lattice one may change from one bond state to another by applying a SU(2) on-site transformation 
$U_A$ and $U_B$ to all of the sites in sublattices A and B, respectively.
Due to the SU (2) invariance of the projector $\mathcal{P}_q$, this is equivalent to performing the transformation to every bond state, 
i.e., $| \omega \rangle \to (U_A)^{1/2} \otimes   (U_B)^{1/2}   | \omega \rangle$.
Thus, given physical data from any bond state, we may produce the corresponding information for another bond state.
Consequently, the location and the nature of the parameter induced transition is same regardless 
the particular $| \omega \rangle$ used on a bipartite lattice. 
This equivalence has been discussed and numerically confirmed in Ref.~\cite{Huang:2016bf, Pomata:2018ep}. 

In this work we focus on the spin-2 AKLT state and the deformed spin-2 AKLT family of states on a 2D square lattice.
Following the recipe above they can easily be constructed by setting $q=4$. Furthermore the deformation matrix reads
$D(\vec{a})=\text{diag}(\frac{a_2}{\sqrt{6}}, \frac{a_1}{\sqrt{6/4}}, 1, \frac{a_1}{\sqrt{6/4}}, \frac{a_2}{\sqrt{6}})$.
This results in a two parameters family of states, while the AKLT point corresponds to $(a_2, a_1, a_0)=(\sqrt{6}, \sqrt{3/2}, 1)$.
It is known that there are parameter induced phase transitions as on tunes one of the $a$s within the parameter space \cite{Niggemann:2000ta, Pomata:2018ep}.
The AKLT point is inside the gapped, disordered phase with SPT order and we will denote this phase as the AKLT-phase.
In the limit of $a_2 \rightarrow \infty$ the state enters a symmetry broken phase 
and become N\'eel or ferromagnetic ordered depending on the particular bond state used.
In this work we use $|\psi^+\rangle$ as the bond state, resulting in a uniform tensor network.
In this case, the symmetry broken phase corresponds to a ferromagnetic (FM) phase with spontaneous uniform magnetization $\langle S^z\rangle$ in the $z$-direction.
It was conjectured in Ref.~\cite{Niggemann:2000ta} that this order-disorder transition corresponds to two simultaneous Ising transitions.
The conjecture is based on the simulation for a system of $30\times 30$ sites. 
The critical exponent $\eta$ is estimated to be $\frac{1}{2}$, which is twice the exact Ising exponent $\frac{1}{4}$.
The transition is further  explored in Ref.\cite{Pomata:2018ep}.
By using TNR, it is found that the central charge $c=\frac{1}{2}$ and the conformal tower obtained by TNR matches the Ising CFT.  

On the other hand, near $a_2=a_1=0$ there is a finite region in which the state is in a XY phase with divergent correlation length.
It is theorized in Ref.\cite{Pomata:2018ep} that the in the continuum limit, the system can be described by the compactified-free-boson CFT,
which has central charge $c=1$. Consequently, the phase transition between the XY phase and the AKLT phase is of BKT type.
In that work, TNR and loop-TNR were used to evaluate the central charge $c$ and the coupling $g$.
The phase boundary is then estimated by locating the position at which the central charge $c$ or the coupling $g$ drops sharply below 1 or 4 respectively.
In Fig.\ref{fig:phase}(a) we sketch the phase diagram based on the known results in the literature.
To study the AKLT-FM transition we fix $a_2=\sqrt{6}$ and vary $a_1$ across the phase boundary.
We also study the limiting case of $a_1=0$ while we vary $a_2$ across the phase boundary.
It is expected that in this limit the universality class is different from 2D Ising model \cite{Pomata:2018ep}.
Finally to study the AKLT-XY transition we fix $a_1=0.5$ and vary $a_2$ across phase boundary.
In Fig.\ref{fig:phase}(a) we also sketch the lines of these three scans.

The spin-2 deformed AKLT states naturally admit a tensor network state (TNS) representation:
\begin{equation}
  |\Psi (\vec{a})  = \sum_{s_1, s_2, \cdots, s_i} 
  \text{tTr} \left(
    A^{1} A^{2} \cdots A^{i} \cdots
  \right) |s_1 s_2 \cdots s_i \cdots \rangle,
\end{equation}
where $A^i_{s_i i_1 i_2 i_3 i_4}$ is a rank-5 local tensor on site-$i$ with a physical index $s_i$ and $4$ virtual bond indices $i_1, i_2,  i_3, i_4 \in [0,1]$.
$\text{tTr}$ denotes tensor trace over all the virtual bond indices. Specifically the non-zero elements of the $A$ tensor are:
\begin{align}
 &  A_{2,1111} =A_{-2,0000} =a_2, \notag \\
 &  A_{1,1110} =A_{1,1101} =A_{1,1011} =A_{1,0111} =a_1,  \notag \\
 &  A_{-1,0001} =A_{-1,0010} =A_{-1,0100} =A_{-1,1000} =a_1,  \notag \\
 &  A_{0,1100}=A_{0,1001}=A_{0,0011}=a_0, \notag \\
 & A_{0,0110}=A_{0,0101}=A_{0,1010}=a_0.
\end{align} 
The norm squared of such a TNS is given by
\begin{equation}
  \langle \Psi | \Psi \rangle = \text{tTr} \left(
    \mathbf{T}^1 \mathbf{T}^2 \cdots \mathbf{T}^i \cdots
  \right)
\end{equation}
where the local {\em doubled tensor} $\mathbf{T}^i$ on site-$i$ is obtained by contracting the physical indices of $A^i$ and $(A^i)^*$:
\begin{equation}
  \mathbf{T}^i_{(i_1 i^\prime_1), (i_2 i^\prime_2), (i_3 i^\prime_3), (i_4 i^\prime_4)} 
  = \sum_{s_i} A^i_{s_i i_1 i_2 i_3 i_4} \times \left(A^i_{s_i i^\prime_1 i^\prime_2 i^\prime_3 i^\prime_4}\right)^*.
\end{equation}
It is convenient to treat double tensor $\mathbf{T}^i$ as a rank-4 tensor, with a compound index $(i i^\prime)\in [0,1,2,3]$ on each leg.
In other word, the bond-dimension for each leg is 4.
It is also straightforward to express the expectation value of operators as a tensor trance. For example,
\begin{equation}
  \langle \Psi | S^{1z} S^{2z} | \Psi \rangle = \text{tTr} \left(
    \mathbf{T}^{1z} \mathbf{T}^{2z} \mathbf{T}^3 \cdots \mathbf{T}^i \cdots
  \right),
\end{equation}
where
\begin{equation}
  \mathbf{T}^{1z} = \sum_{s_i} A^i_{s_i i_1 i_2 i_3 i_4} \times \hat{S}^{1z} \times \left(A^i_{s_i i^\prime_1 i^\prime_2 i^\prime_3 i^\prime_4}\right)^*
\end{equation}
and similarly for $\mathbf{T}^{2z}$.
Due to the spin-flip symmetry, in the FM phase $|\Psi (\vec{a})\rangle$ is a superposition of both possible ordered states, and the magnetization is strictly zero. 
One can apply a very tiny symmetry breaking field to induce a non-zero magnetization,
but one has numerically take the zero field limit to obtain the spontaneous magnetization.
In this work, we will use the second and fourth moments of the magnetization to characterize the ordered phase.
These even moments can pick up non-zero value in the absence of the external field.

In general, performing exactly the tensor trace in two and higher dimension is exponentially difficult.
There are, however, many approximation schemes which scale down the cost to the polynomial of cut-off bond dimension.
These include, for example, corrner transfer matrix (CTMRG) \cite{Nishino:1997kn, Orus:2012ft}, tensor renormalization group (TRG) \cite{Levin:2007jua}, 
higher-order tensor renormalization group (HOTRG) \cite{Xie:2012iy}, etc. 
In this work we mainly use HOTRG. 
While HOTRG is often used to study the system in thermodynamic limit,
here we focus on the finite-size system with linear size $L=2^{N+1}$, where $N$ is number of HOTRG steps.
We note in passing that the accuracy of the HOTRG is determined by the cut-off dimension $D_{\text{cut}}$.
In the following we set $D_{\text{cut}}=50$ unless mentioned otherwise.
We have checked that this cut-off dimension is large enough for the calculations in this work.
Furthermore, we use the method proposed in Ref.~\cite{Morita:2018gy} to evaluate the higher moments at different sizes.

\section{AKLT-FM transition}
\label{AKLT-FM}

In this section we study the AKLT-FM transition and demonstrate that 
the critical point and the critical exponents can be determined accurately using tensor network methods 
based FSS analysis.
We start from the AKLT point and vary $a_1$ across the phase boundary as shown in Fig.~\ref{fig:phase}(a).
Two approaches are used to estimate the critical point and the critical exponents. 
Both approaches rely on the finite-size scaling hypothesis, which states that near a continuous phase transition a quantity $Q$ shall scale as
\begin{equation}
  Q(a, L) = L^{c_2} f( (a-a_c)L^{c_1}),
\end{equation}
where $a$ is the tuning parameter, $a_c$ is the critical point, $c_1, c_2$ are the critical exponents, and $f$ is the scaling function.
In the first approach, the critical point $a_c$ and the critical exponents $c_1, c_2$ are estimated {\em simultaneously} 
by collapsing the data of Q from various $a$ and $L$. 
In this work, we use the the kernel method proposed in Ref.\cite{Harada:2011js} to perform the scaling analysis.
In the second approach we first use the crossing point of the dimensionless quantities to determine the critical point,
then we use the finite-size scaling at the critical point to estimate the critical exponents. 
Since a dimensionless quantity $\tilde{Q}$ shall scale as
\begin{equation}
  \tilde{Q}(a, L) = f( (a-a_c) L^{c_1}),
\end{equation}
one finds $\tilde{Q}(a_c, L)=f(0)$ and data from different sizes should cross at the critical point $a_c$. 
However, if the correction to the scaling is not negligible the crossing point will drift as $L$ increases.  
In this case a better estimation of the critical point can be obtained by extrapolating the crossing points.

Consider the $k$-th moment of the uniform magnetization in the $x, y, z$ directions:
$\langle m^k_{x, y, z} \rangle \equiv \langle (\frac{1}{L^2}\sum_{i} S^{ x, y, z}_i)^k \rangle$.
These moments can be calculated via HOTRG by using the procedure proposed in Ref.\cite{Morita:2018gy}.
Since there is no magnetic order in the AKLT phase one has $\langle m^k_{x,y,z} \rangle=0$. 
On the other hand in the FM phase one has $\langle m^k_z \rangle \neq 0$ and $\langle m^k_{x,y}\rangle=0$.
Near the phase transition the $k$-th moment $\langle m^k_z\rangle$ shall scale as  
\begin{equation}
  \langle m^k_z \rangle (a, L) = L^{-k\beta/\nu} f_k( (a-a_c) L^{1/\nu}),
\end{equation}
where $\beta$ is the standard critical exponent associated with the magnetization $ m \propto (a-a_c)^\beta$ and $f_k$ is the scaling function. 
In particular, we calculate the second moment $\langle m^2\rangle$ and the fourth moments $\langle m^4\rangle$.
Furthermore, we consider the Binder ratio of the fourth moment and the square of second moment, $U_2 \equiv \langle m^4\rangle/ \langle m^2\rangle^2$.
Since it is dimensionless it shall scale as
\begin{equation}
  U_2(a, L) = \frac{\langle m^4\rangle}{\langle m^2\rangle^2} = f_U( (a-a_c) L^{1/\nu}).
\end{equation}

\begin{figure}[t]
  \includegraphics[width=1.0\columnwidth]{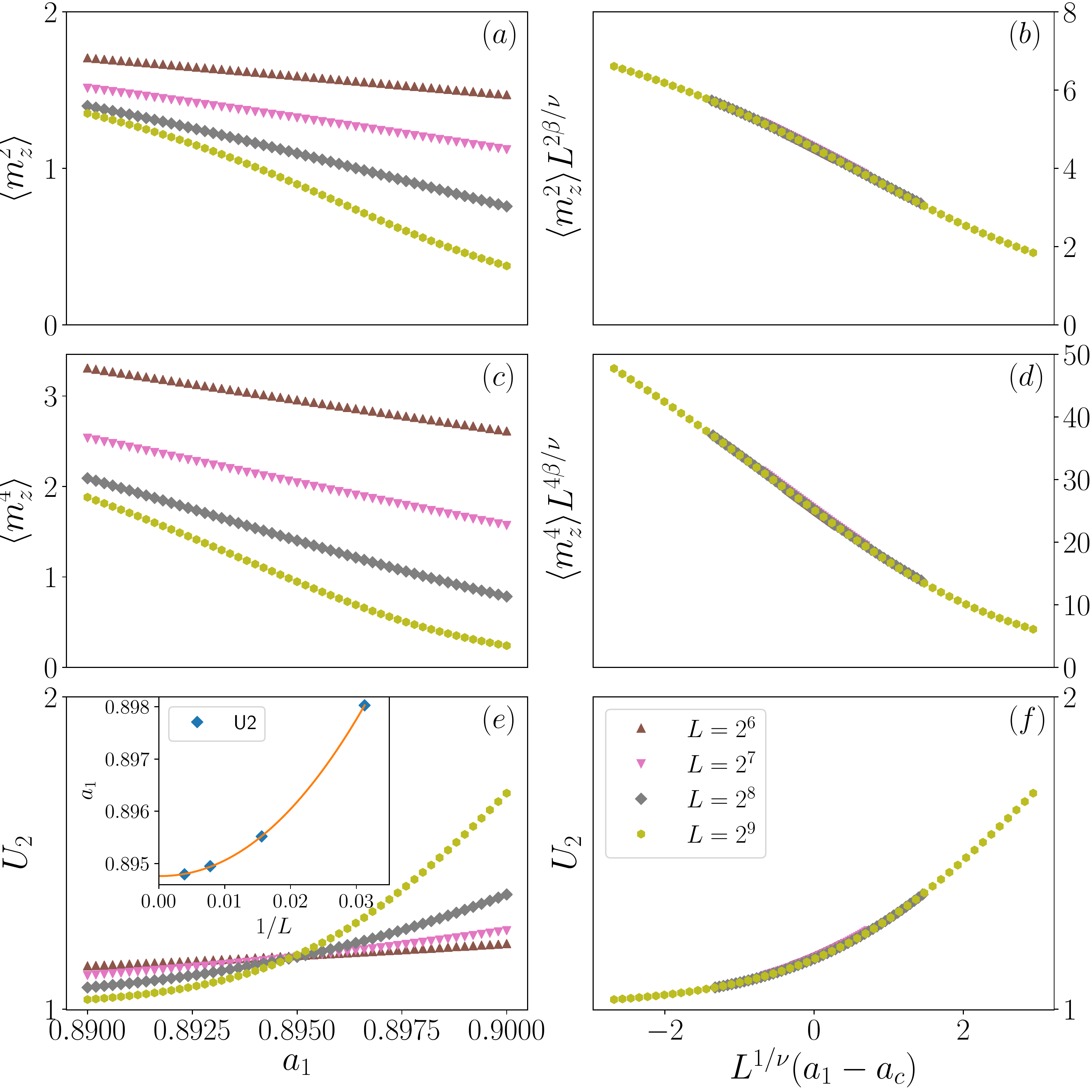}  
  \caption{(Color online) (a) Second moment $\langle m^2_z \rangle$, 
  (c) forth moment $\langle m^4_z \rangle$, (e) Binder ratio $U_2$ as a function of $a_1$ for $L=64, 128, 256, 512$. 
  (b), (d), (f) Rescaled $\langle m^2_z \rangle$, $\langle m^4_z \rangle$, and $U_2$ as a function of $ (a_1-a_c)  L^{1/\nu}$. 
  Inset of (e)  Crossing points $a_{c}(L)$ of $U_2(L)$ and $U_2(2L)$ as a function of $1/L$.} 
  \label{fig:BSA}
\end{figure}

\begin{figure}[t]
  \includegraphics[width=1.0\columnwidth]{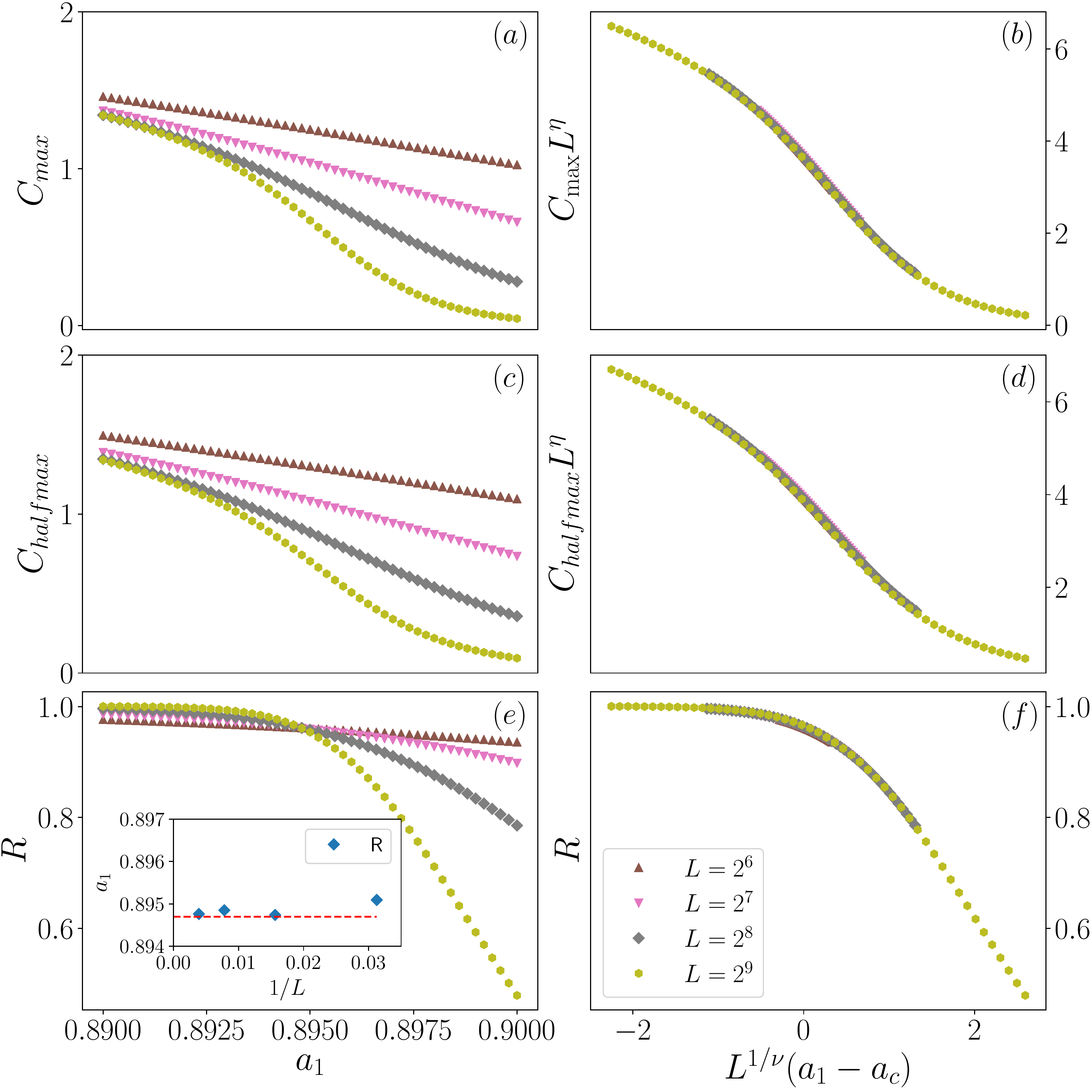}  
  \caption{(Color online) (a) $C_{\text{max}}(L)$, (c) $C_{\text{halfmax}}(L)$, (e) correlation ratio $R$ as a function of $a_1$ for $L=64, 128, 256, 512$. 
  (b),(d),(f) Rescaled $C_{\text{max}}(L)$, $C_{\text{halfmax}}(L)$, and $R$ as a function of $ (a_1-a_c) L^{1/\nu}$.
  Inset of (e)  Crossing points $a_{c}(L)$ of $R(L)$ and $R(2L)$ as a function of $1/L$.}
  \label{fig:BSA_C}
\end{figure}

In Fig.\ref{fig:phase}(b) we plot the second moment $\langle m^2_z\rangle$ near the AKLT-FM transition.
We fix $a_2=\sqrt{6}$ and vary $a_1$ across the phase boundary. 
We observe that the second moment become non-zero around $a_1 \approx 0.9$.
Furthermore, the transition become sharper as system size increases, indicating a second-order phase transition in the thermodynamic limit.
In Fig.\ref{fig:BSA}(a), (c), (e) we show $\langle m^2_z \rangle$, $\langle m^4_z \rangle$, and Binder ratio $U_2$ near the critical point.
By using  the kernel method to collapse the data from different sizes we estimate the critical point and critical exponents.
In Fig.\ref{fig:BSA}(b),(d),(f) we show the rescaled data and we observe that the data collapse very well.
For the critical point, we find $a_c \approx 0.894(6), 0.894(7), 0.894(8)$ respectively from $\langle m^2_z \rangle$, $\langle m^4_z \rangle$, and $U_2$.
We also find $1/\nu \approx 1.0(1), 1.0(1), 1.0(2)$, where $\nu$ is the exponent associated with the correlation length $\xi \propto (a-a_c)^{-\nu}$.
Furthermore we find $2\beta/\nu \approx 0.2(6)$ from $\langle m^2_z \rangle$ and $4\beta/\nu \approx 0.5(2)$ from $\langle m^4_z \rangle$.
The estimated critical points obtained with different quantities are very close to each other,
and the values of the exponent are consistent with the expected 2D Ising universality class.

Next we consider the spin-spin correlation function in the $x$, $y$, and $z$ directions,
\begin{equation}
  C^{x,y,z}(\mathbf{r}) \equiv \frac{1}{L^2} \sum_{\mathbf{r}_i} \langle S^{x,y,z}_{\mathbf{r}_i} S^{x,y,z}_{\mathbf{r}_i + \mathbf{r}}\rangle.
\end{equation}
For finite system near the critical point, there are two length scales: system size $L$ and correlation length $\xi$.
On general ground the following scaling form with two scaling variables is expected:
\begin{equation}
  C^{x,y,z}(\mathbf{r}) = |\mathbf{r}|^{-(D-2+\eta)} h_{x,y,z}(\mathbf{r}/L, L/\xi),
\end{equation}
where $\eta$ is the anomalous dimension and $h_{x,y,z}$ are the scaling functions.
In particular we calculate the correlation at maximum distance: $C_{\text{max}}(L) \equiv C^{x,y,z}((L/2,L/2))$ and 
half of maximum distance: $C_{\text{halfmax}}(L) \equiv C^{x,y,z}((L/4,L/4))$.
In passing we note that the correlation at maximum and half-maximum distance can be calculated efficiently using HOTRG.
In a conventional second-order phase transition where the correlation length diverges as a power law $\xi \propto t^{-\nu}$, one has
\begin{equation}
  C_{\text{max, halfmax}}(a, L) = L^{-(D-2+\eta)} h_{\text{max, halfmax}}( (a-a_c) L^{1/\nu}),
\end{equation}
where $h_{\text{max, halfmax}}$ are scaling functions and $D$ is the dimension of the system.
In addition, we consider the dimensionless correlation ratio $R$ of $C_{\text{max}}(L)$ and $C_{\text{halfmax}}(L)$, which should scale as
\begin{equation}
  R(a, L) \equiv \frac{ C_{\text{max}}(a,L) }{C_{\text{halfmax}}(a,L) } = h_R( tL^{1/\nu} ),
\end{equation}
with some scaling function $h_R$.

In Fig.\ref{fig:BSA_C}(a), (c), (e) we show $C_{\text{max}}(L)$, $C_{\text{halfmax}}(L)$, and correlation ratio $R$ as a function of $a_1$, 
while in Fig.\ref{fig:BSA_C}(b), (d), (f) we show the rescaled data. 
By collapsing the data we find $a_c \approx 0.894(7), 0.894(7), 0.894(8)$ and 
$1/\nu \approx 1.0(1), 1.0(1), 1.0(4)$  respectively from $C_{\text{max}}(L)$, $C_{\text{halfmax}}(L)$, and $R$.
We observe again that the estimated critical points are highly consistent with each other as well as the results from moments and Binder ratio.
Furthermore, from both $C_{\text{max}}(L)$ and  $C_{\text{halfmax}}(L)$ we find $\eta \approx 0.2(5)$.
These values are again consistent with the 2D Ising universality class.

\begin{figure}[t]
  \includegraphics[width=1.0\columnwidth]{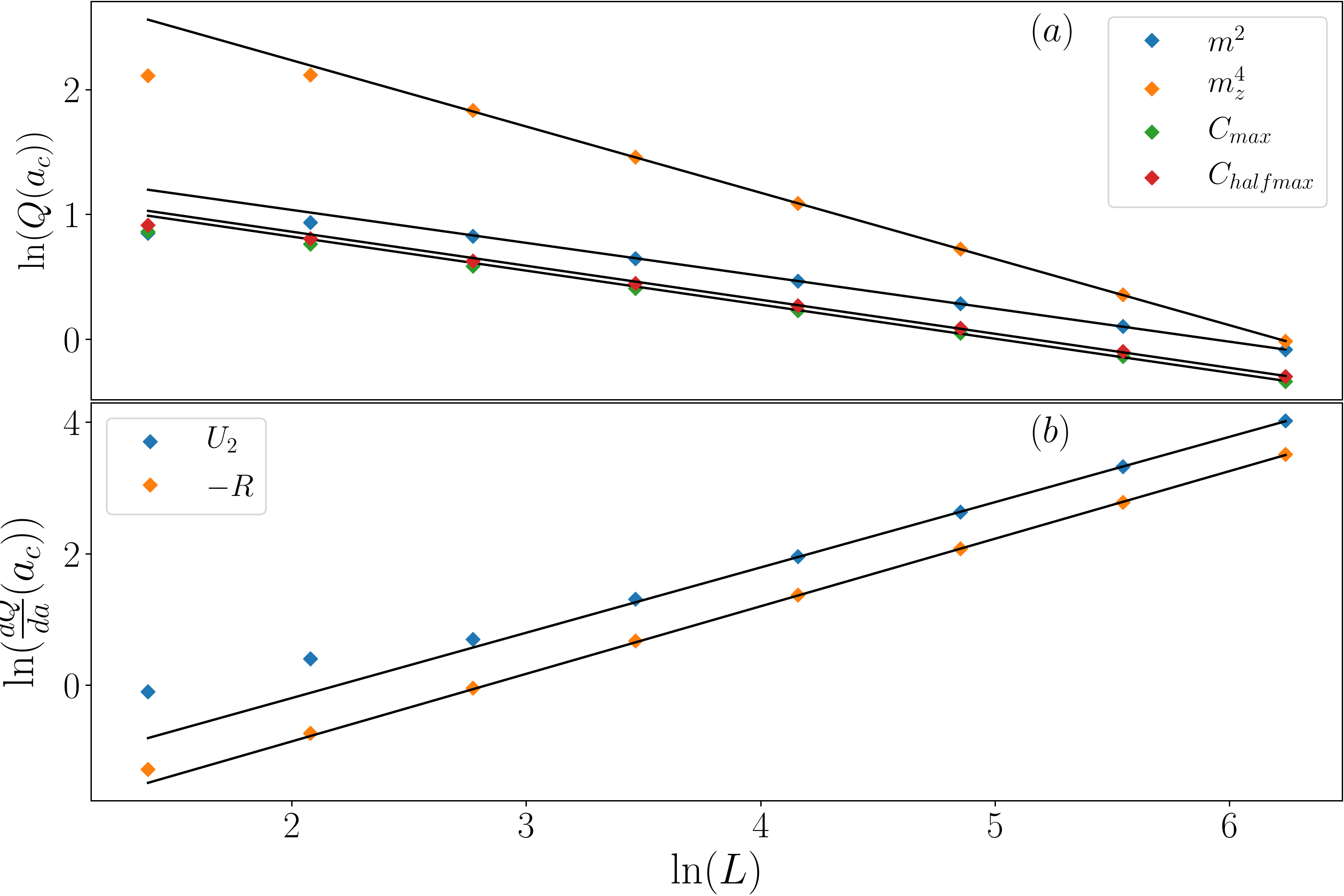}  
  \caption{(Color online) (a) $\ln(Q)$ as a function of $\ln(L)$, where $Q=\langle m^2_z\rangle, \langle m^4_z\rangle, C_{\text{max}}, C_{\text{halfmax}}$. 
  Finite-size scaling at critical point. (b) $\ln\left(\frac{dQ}{da}\right)$ as a function of $\ln(L)$, where $Q=U_2$ and $-R$.}
  \label{fig:Q_ac}
\end{figure}

Now we move on to the second approach. 
In this approach, we first use the crossing point of the dimensionless quantities $U_2$ and $R$ to estimate the critical point.
We then study the finite-size scaling of various quantities at the critical point to estimate the corresponding exponents.
In the inset of Fig.\ref{fig:BSA}(e) we plot the crossing points $a_{c,U_2}(L)$ of $U_2(L)$ and $U_2(2L)$ as a function of $1/L$. 
We observe that $a_{c,U_2}(L)$ decreases monotonically as $L$ increases. 
By fitting the finite-size crossing points to a power-law function $a_{c,U_2}(L) = a_c + b L^\lambda$ we find $a_c \approx 0.894(8)$.
In the inset of Fig.\ref{fig:BSA_C}(e) we plot the crossing points $a_{c,R}(L)$ of $R$ as a function of $1/L$.
In this case we find that the crossing points of $R$ does not drift much and we estimate $a_c \approx 0.894(8)$ by the crossing point of the largest size.
In passing we note that these results are highly consistent with the results from the first approach.

After locating the critical point, the values of the critical exponents can be estimated by studying the finite-size scaling of various quantities at the (estimated) critical point. At the critical point the $k$-th moment shall scales as
\begin{equation}
  \langle m^k_z \rangle(a=a_c, L) \propto L^{k\beta/\nu},
\end{equation}
while the correlation function shall scale as
\begin{equation}
  C_{\text{max, halfmax}}(a=a_c, L) \propto L^{D-2+\eta}.
\end{equation}
Finally the critical exponent $\nu$ can also be estimated by the derivates of the dimensionless quantities at the critical point
\begin{equation}
  \left. \frac{dU_2(a,L)}{da} \right|_{a_c} \propto L^{1/\nu},
\end{equation}
and similarly for $-dR/da$. Here we obtain the slope via numerical differentiation.

In Fig.~\ref{fig:Q_ac} we plot the log of above mentioned quantities as a function of $\ln(L)$.
We use data from $L=64,128,256,512$ to perform the liner fit and the slop of the fitted line is the corresponding exponent.
In the figure, we also show data from smaller sizes with $L=4,8,16,32$ and deviation from the liner fit is clearly observed.
From the fitting we find $2\beta/\nu \approx 0.2(6)$ from $\langle m^2_z\rangle$ and $4\beta/\nu \approx 0.5(3)$ from $\langle m^4_z\rangle$.
For the exponent $\eta$, we find $\eta \approx 0.2(7)$ from both $C_{\text{max}}(L)$ and $C_{\text{halfmax}}(L)$.
Finally from $dU_2/da$ and $-dR/da$ we find $1/\nu \approx 0.9(9)$ and $1.0(3)$. 
All these values are consistent with the 2D Ising universality class. 

In Table.~\ref{tb:AKLT-FM_ac} we summarize the values of the estimated critical point.
The label $X$ indicates that the results are obtained from the crossing point analysis.
In Table.~\ref{tb:AKLT-FM} we summarize the values of the exact and estimated critical exponents.
The label $a_c$ indicates that the results are obtained from the finite-size scaling at the critical point.
We observe that for both methods used in this section, the estimated critical point are highly consistent with each other.
Furthermore, the estimated critical exponents are highly consistent with the expected 2D Ising universality class.
This demonstrates that the tensor network based FSS analysis can be used 
to determine precisely the critical point as well as the critical exponents for a second-order phase transition.


It was pointed out in Ref.\cite{Pomata:2018ep} that as $a_1 \rightarrow 0$ the central charge $c$ becomes $1$.
As a result, in this limit the AKLT-FM transition can not belong to the 2D Ising universality class.
To study the phase transition in this limit we fix $a_1=0$ and vary $a_2$ across the phase boundary.
We then apply the above mentioned finite-size scaling analysis to estimate the critical point and exponents.
The detail of the finite-size scaling analysis is presented in appendix~\ref{a1=0}.
From the data collapse and the crossing points of $U_2$ and $R$ 
we find  $a_c \approx {1.779(6)}$, which is consistent with the result in Ref.\cite{Pomata:2018ep}.
Furthermore, the exponents obtained are clearly different from the exponents of the 2D Ising model. 
This confirms that the transition at $a_1=0$ does not belong to the 2D Ising universality class.

\begin{widetext}

\begin{table}[tb]
  \caption{Summary of the estimated critical point. ($X$ indicates that the results are obtained from the crossing point analysis.)}
  \label{tb:AKLT-FM_ac}
  \begin{center}
  \begin{tabular}{|c|c|c|c|c|c|c|c|c|}
  \hline \hline  
          & $m^2_z$ & $m^4_z$ & $U_2$  & $C_{\text{max}}$ & $C_{\text{halfmax}}$ & $R$ & $U_2$(X) & $R$(X) \\  \hline
   $a_c$ & {0.894(6)} & {0.894(7)} & {0.894(8)}  & {0.894(7)} & {0.894(7)} & {0.894(8)}  & {0.894(8)} & {0.894(8)} \\  \hline
  \end{tabular}
  \end{center}
\end{table}

\begin{table}[tb]
  \caption{Summary of the exact and estimated critical exponents.}
  \label{tb:AKLT-FM}
  \begin{center}
  \begin{tabular}{|c|c|c|c|c|c|c|c|c|c|c|c|c|c|}
  \hline \hline  
  &  2D Ising & $m^2_z$  & $m^2_z(a_c)$ & $m^4_z$ &  $m^4_z(a_c)$ & $U_2$ & $C_{\text{max}}$ & $C_{\text{max}}(a_c)$ & $C_{\text{halfmax}}$ &  $C_{\text{halfmax}}(a_c)$ & $R$ & $\left.\frac{dU_2}{da}\right|_{a_c}$ & $\left.\frac{dR}{da}\right|_{a_c}$ \\ \hline
  $1/\nu$ & 1  & {1.0(1)} & - & {1.0(1)} & - & {1.0(2)} & {1.0(1)}  & -& {1.0(1)} & - & {1.0(4)} & {0.9(9)} & {1.0(3)} \\   \hline
  $2\beta/\nu$ & 1/4  & {0.2(6)} & {0.2(6))} & - & - & - & - & - & - & - & - & - & -\\   \hline
  $4\beta/\nu$ & 1/2  & - & - & {0.5(2)} & {0.5(3)} & - & - & - & - & - & - & - & -\\   \hline  
  $\eta$ & 1/4  & - & - & - & - & - & {0.2(5)} & {0.2(7)} & {0.2(5)} & {0.2(7)} & - & - & - \\   \hline    
  \end{tabular}
  \end{center}
\end{table}

\end{widetext}
\section{AKLT-XY transition}
\label{AKLT-XY}

\begin{figure}[t]
  \includegraphics[width=1.0\columnwidth]{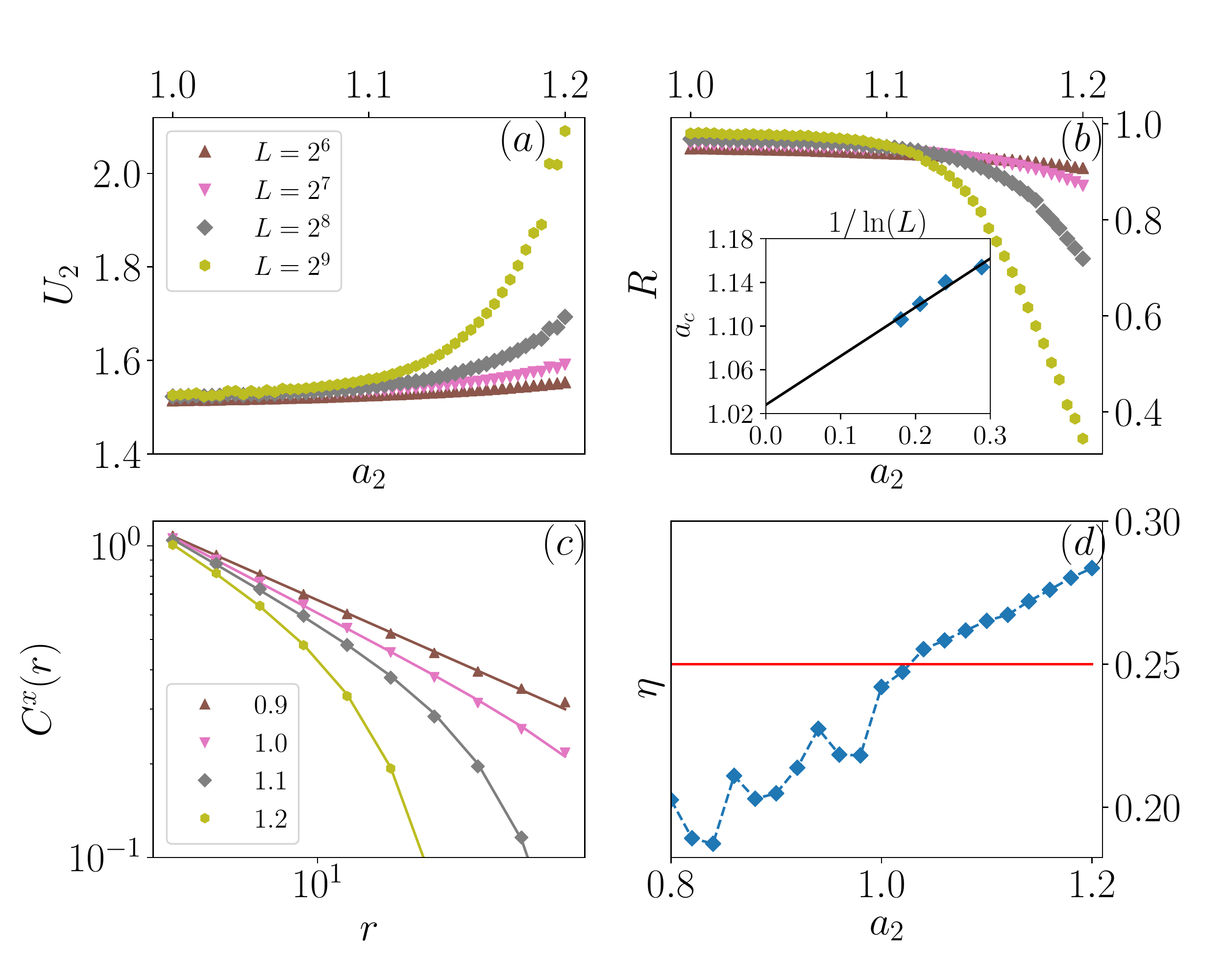}  
  \caption{(Color online) (a) Binder ratio $U_2$ as a function of $a_2$. 
  (b) Correlation ratio $R$ as a function of $a_2$. Inset: Crossing point $a_c$ of $R(L)$ and $R(2L)$ as a function of $1/\ln(L)$.
  (c) Correlation function $C^x(r)$ with various $a_2$.
  (d) Fitted $\eta$ as a function of $a_2$.}
  \label{fig:AKLT-XY}
\end{figure}

In this section we study the AKLT-XY transition. In Ref.~\cite{Pomata:2018ep} it is theorized that in the critical XY phase
the system is described by the compactified-free-boson CFT with central charge $c=1$ and 
the phase transition between the XY phase and the AKLT phase is of BKT type. 
In that work, TNR and loop-TNR were used to evaluate the central charge $c$ and the coupling $g$.
The phase boundary is estimated by locating the position at which the central charge $c$ or the coupling $g$ drops sharply below 1 or 4 respectively.
In this work, we will use tensor network based finite-size scaling analysis to investigate the AKLT-XY transition.
In general it is difficult to accurately determine the phase boundary of a BKT transition.
Inside the XY phase the correlation length diverges in the thermodynamic limit,
while it grows exponentially as one approaches the XY phase.
In this case the correlation ratio reads
\begin{equation}
  R(a, L) \equiv \frac{ C_{\text{max}}(a,L) }{C_{\text{halfmax}}(a,L) } = h_R( L/\xi(a) ).
\end{equation}
Due to the divergence of the correlation length, the correlation ratio in the XY phase becomes $h_R(0)$.
Consequently data from different sizes should collapse within the XY phase, in contrast to cross at the critical point for a second-order phase transition.
In Ref.~\cite{Tomita:2002cka, Surungan:2019hjf} it is proposed and tested that correlation ratio is better than Binder ratio in locating the BKT transition,
due to the cancelation of the logarithmic corrections \cite{Nomura:1995tq}.

In the following we preset our results for the AKLT-XY transition.
Specifically, we fix $a_1=0.5$ and vary $a_2$ across the phase boundary.
Similar to the study of the AKLT-FM transition, we evaluate the second and forth moments, 
the correlation functions at maximum and half-maximum distance as well as the dimensionless Binder ratio and correlation ratio.
In Fig.\ref{fig:phase}(c) we plot the second moment of the magnetization in the $x$-direction $\langle m^2_x\rangle$ as a function of $a_2$.
We observe that as $a_2$ decreases $\langle m^2_x\rangle$ becomes non-zero. 
However, for a fixed $a_2$ the value of $\langle m^2_x\rangle$ decreases monotonically as system size increases, 
indicating that in the thermodynamics limit one has $\langle m^2_x\rangle=0$.
Furthermore, we have checked that $\langle m^2_x\rangle = \langle m^y_2\rangle$ and $\langle m^2_z\rangle=0$ within this parameter regime.
Qualitatively, these results are consistent with a BKT transition.

In Fig.~\ref{fig:AKLT-XY}(a) we plot the Binder ratio $U_2$ as a function of $a_2$ near the phase boundary.
We find that curves from different sizes never cross. Furthermore, they do not merge in the XY phase either.
As a result, one cannot use Binder ratio to locate the BKT transition point.
In Fig.~\ref{fig:AKLT-XY}(b) we plot the correlation ratio $R$ as a function of $a_2$. In this case the curves from different sizes do cross each other.
However, they do not collapse in the XY phase. Similar phenomenon was observed in Ref.~\cite{Komura:2010hz}, 
in which the correlation ratio is used to study the BKT transition of the generalized XY model.
The non-merging suggests that for this BKT transition, the correction to FSS is extremely large.
To better estimate the critical point, we plot the crossing point $a_{c,R}(L)$ of $R(L)$ and $R(2L)$ 
as a function of $1/\ln(L)$ in the inset of Fig.~\ref{fig:AKLT-XY}(b).
We observe that all data fall nearly on a straight line. 
By linear fitting we find $a_c \approx 1.0(3)$, which is consistent with the phase boundary estimated in Ref.~\cite{Pomata:2018ep}.

We also evaluate the spin-spin correlation function $C^x(r) = \langle S^x(0,0) S^x(r,r)\rangle$ near the phase boundary,
as shown in Fig.~\ref{fig:AKLT-XY}(c).
In the thermodynamic limit the spin-spin correlation function should decay as a power law $ r^{-\eta}$ in the XY phase,
while it should decay as $ r^{-\eta} e^{-r/\xi}$ in the AKLT phase.
Furthermore, it is hypothesized in Ref.~\cite{Pomata:2018ep} that $\eta=1/4$ at the AKLT-XY phase boundary.
To better estimate $\eta$, we study a system with linear size $L=2048$ with a larger cut-off bond-dimension in HOTRG, $\text{D}_{\text{cut}}=56$.
Without assuming the precise location of the critical point, we fit data to both decay forms mentioned above.
We find that the second form always fits the data well.
In contract, the first form fits well only when the system is sufficiently inside the XY phase.
However, when the first form fits well, the fitted value of $\eta$ is very close to one obtained by second form.
In Fig.~\ref{fig:AKLT-XY}(d) we plot $\eta$ obtained by second form as a function of $a_2$. 
We observe that $\eta$ decrease almost linearly as it approaches the XY phase.
Furthermore, we find that $\eta \approx 1/4$ at the estimated $a_c$, consistent with the finding in Ref.~\cite{Pomata:2018ep}.

\section{Summary and Discussion\label{discussion}}

In this work we present a tensor network based finite-size scaling analysis 
on the parameter induced phase transitions of the deformed AKLT family of states on 2D square lattice. 
In particular we use HOTRG to evaluate the moments, the correlation functions, and their dimensionless ratios on a finite-size system. 
We then use conventional FSS techniques to estimate the critical points and exponents.
Two approaches are used. In the first approach, we estimate the critical point and exponent simultaneously via data collapse.
In the second approach the critical point is first located by the crossing point of the dimensionless quantities.
The finite-size scaling of various quantities at the estimated critical point is then used to estimate the corresponding exponents.
For the AKLT-FM transition, we show that both the critical point and critical exponents can be estimated accurately.
Furthermore, the estimated critical exponents are highly consistent with the expected 2D Ising universality class.
We also study the limiting case of $a_1=0$ and confirm that the AKLT-FM transition in this limit is not Ising-like.
For the more elusive BKT type AKLT-XY transition, we demonstrate that the crossing point of the correlation ratio can be used
to locate the critical point with reasonable accuracy. We also estimate the critical exponent $\eta$ by fitting the spin-spin
correlation function and show that $\eta$ becomes $1/4$ at the phase boundary as expected by the theory proposed in Ref.\cite{Pomata:2018ep}.

Some comments are now in order. Typically the tensor network method is used to evaluate quantities in the thermodynamics limit,
where spontaneous symmetry breaking can happen. However, contracting exactly the 2D tensor network is exponentially difficult,
and certain cut-off on the bond-dimension has to be implemented to keep the calculation manageable.
This effectively induces an upper limit on the correlation length . 
Consequently near the critical point, where the correlation length diverges, the tensor network method 
may fail to capture the true critical behavior but show mean-field like behavior \cite{Liu:2010dl}. 
In this work we take a different approach. 
We use tensor network methods to evaluate physical quantities in a finite size system.
We look for system sizes, which are large enough to show scaling behavior, 
but small enough so that physical quantities evaluated via HOTRG are accurate enough.
We then employ conventional FSS analysis to extract the critical point and the critical exponents.
Our results show that this approach is feasible, opening new directions in using tensor network method to study critical properties.

To the best of knowledge, tensor network based FSS analysis is not widely used in the literature.
One of the reason is that an efficient algorithm to calculate higher-order moments for 2D tensor network is not proposed until recently \cite{Morita:2018gy},
while higher-order moments and their dimensionless ratio are essential quantities to be used in conventional FSS analysis.
By using the above mentioned algorithm, we are able to perform FSS analysis based on moments and accurate results are obtained.
Furthermore, we also demonstrate that one can perform FSS analysis based on correlations at different distances and their dimensionless ratio.
Since it is relatively straightforward to evaluate correlations using tensor network methods, 
it is very interesting to explore potential applications in this direction.

Finally, we would like to point out potential generalizations in the future.
Recently, several new renormalization schemes of tensor networks are proposed.
For example, the core-tensor RG \cite{Lan:2019et},
the anisotropic TRG \cite{Adachi:2019paf}, and the triad RGs \cite{Kadoh:2019tp}.
The main motivation is to reduce the scaling of the algorithm in terms of the cut-off and physical dimensions,
and higher dimensional calculations become feasible.
It would be very interesting to use these schemes to perform tensor network based FSS on higher-dimensional systems.
Furthermore, due to the nature of the HOTRG, it is cumbersome to reach systems with linear sizes that are not a power of two.
On the other hand, method such as core-tensor RG can reach any system size with the same effort.
We expect that an even better FSS analysis can be achieved by accessing more sizes in the scaling regime.

\begin{acknowledgments}
This work was supported by the MOST of Taiwan under Grants No. 107-2112-M-007-018-MY3 and No. MOST108-2112-M-029-006-MY3.
Pochung Chen thanks Kenji Harada for helpful discussions in using his Bayesian scaling analysis toolkit.
The numerical calculation was done using the Uni10 tensor network library \cite{Kao:2015gb}. https://uni10.gitlab.io/.
\end{acknowledgments}

\appendix

\section{AKLT-FM transition at $a_1=0$}
\label{a1=0}

\begin{figure}[t]
  \includegraphics[width=1.0\columnwidth]{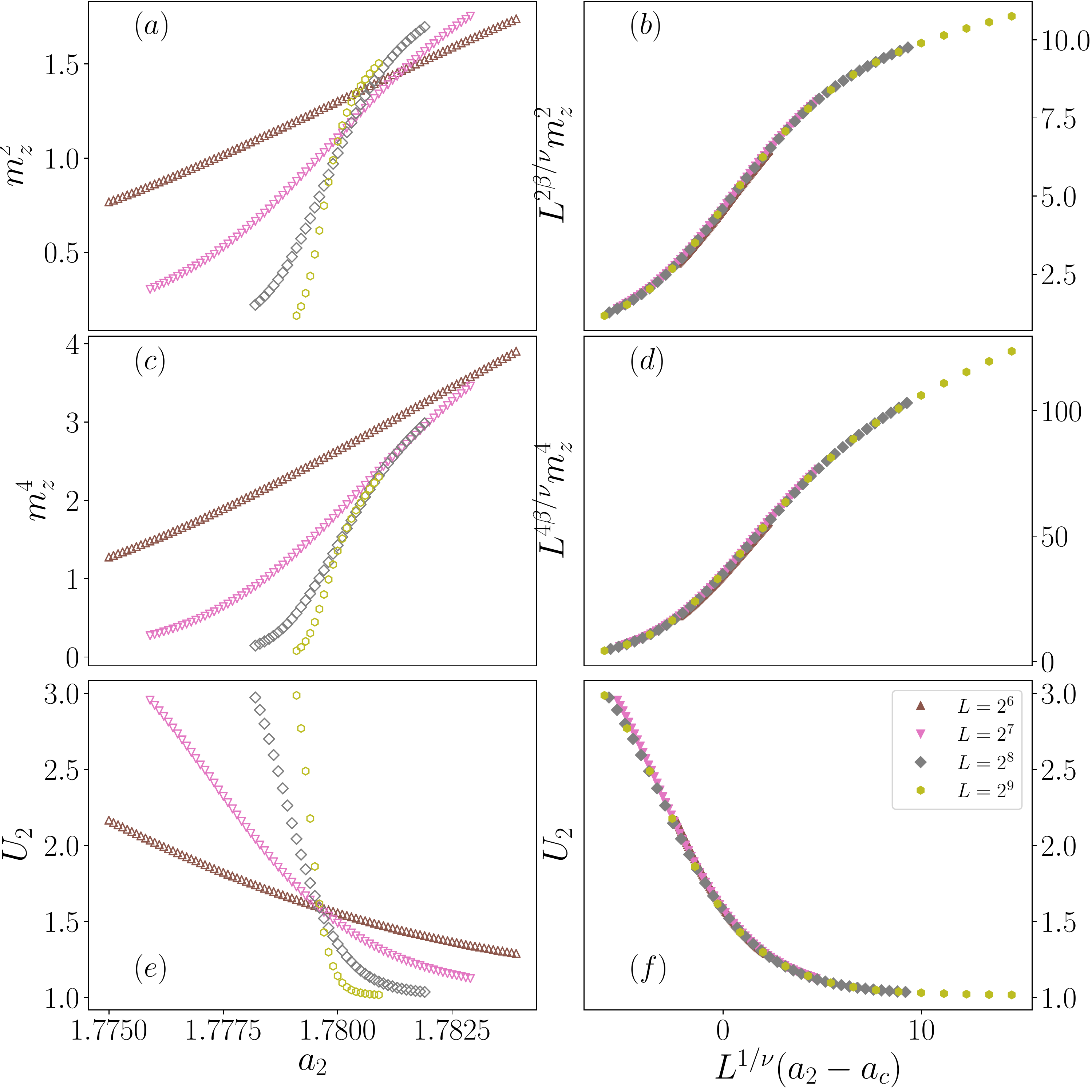}  
  \caption{(Color online) (a) Second moment $\langle m^2_z \rangle$, (c) forth moment $\langle m^4_z \rangle$, (e) Binder ratio $U_2$ as a function of $a_1$ for $L = 64, 128, 256, 512$. 
  (b), (d), (f) Rescaled $\langle m^2_z \rangle$, $\langle m^4_z \rangle$, and $U_2$ as a function of $ (a_1-a_c)  L^{1/\nu}$.}
  \label{fig:BSA_a0}
\end{figure}

\begin{figure}[t]
  \includegraphics[width=1.0\columnwidth]{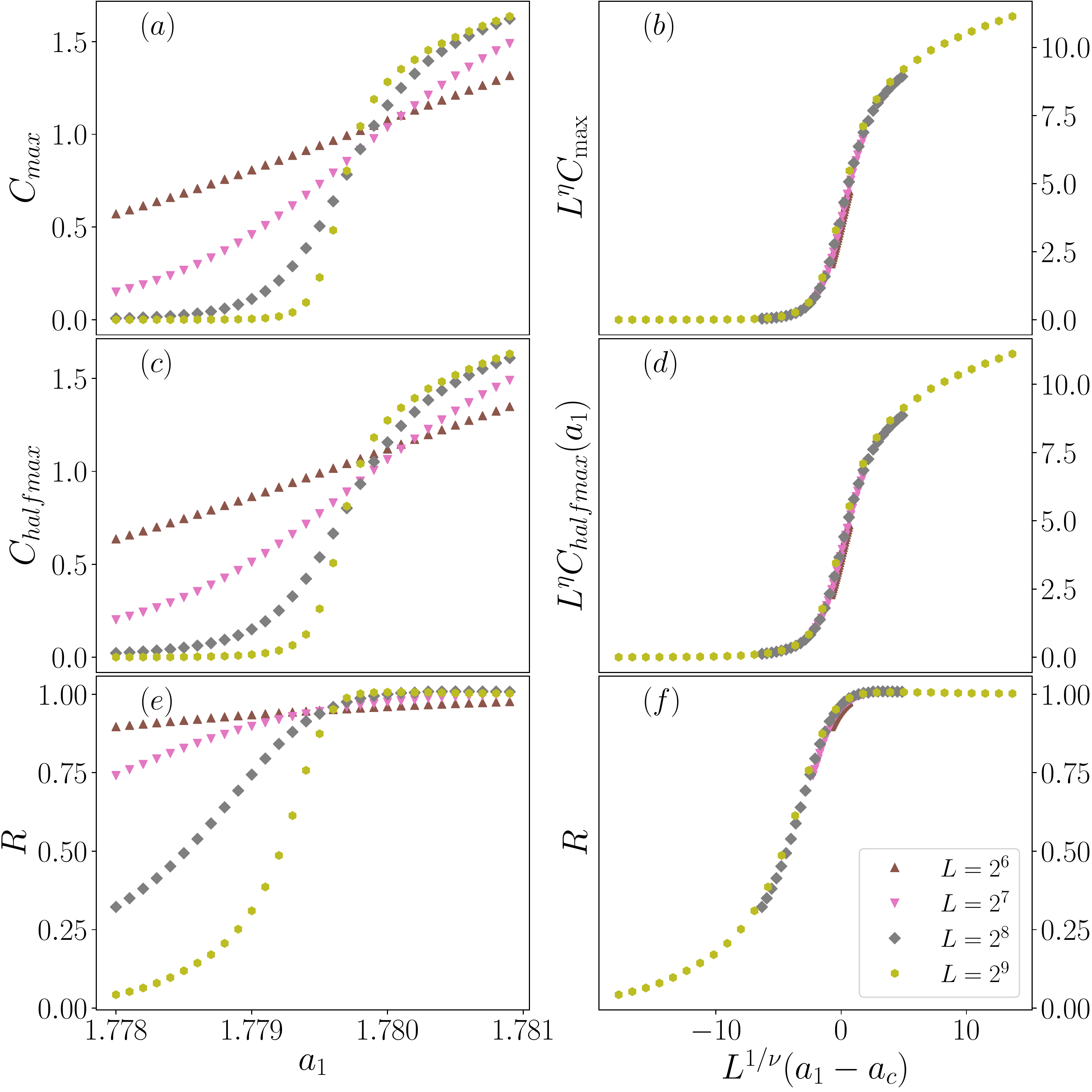}  
  \caption{(Color online) (a) $C_{\text{max}}(L)$, (c) $C_{\text{halfmax}}(L)$, (e) correlation ratio $R$ as a function of $a_1$ for $L = 64, 128, 256, 512$. 
  (b),(d),(f) Rescaled $C_{\text{max}}(L)$, $C_{\text{halfmax}}(L)$, and $R$ as a function of $ (a_1-a_c) L^{1/\nu}$.}
  \label{fig:BSA_C_a0}
\end{figure}

\begin{figure}[h]
  \includegraphics[width=1.0\columnwidth]{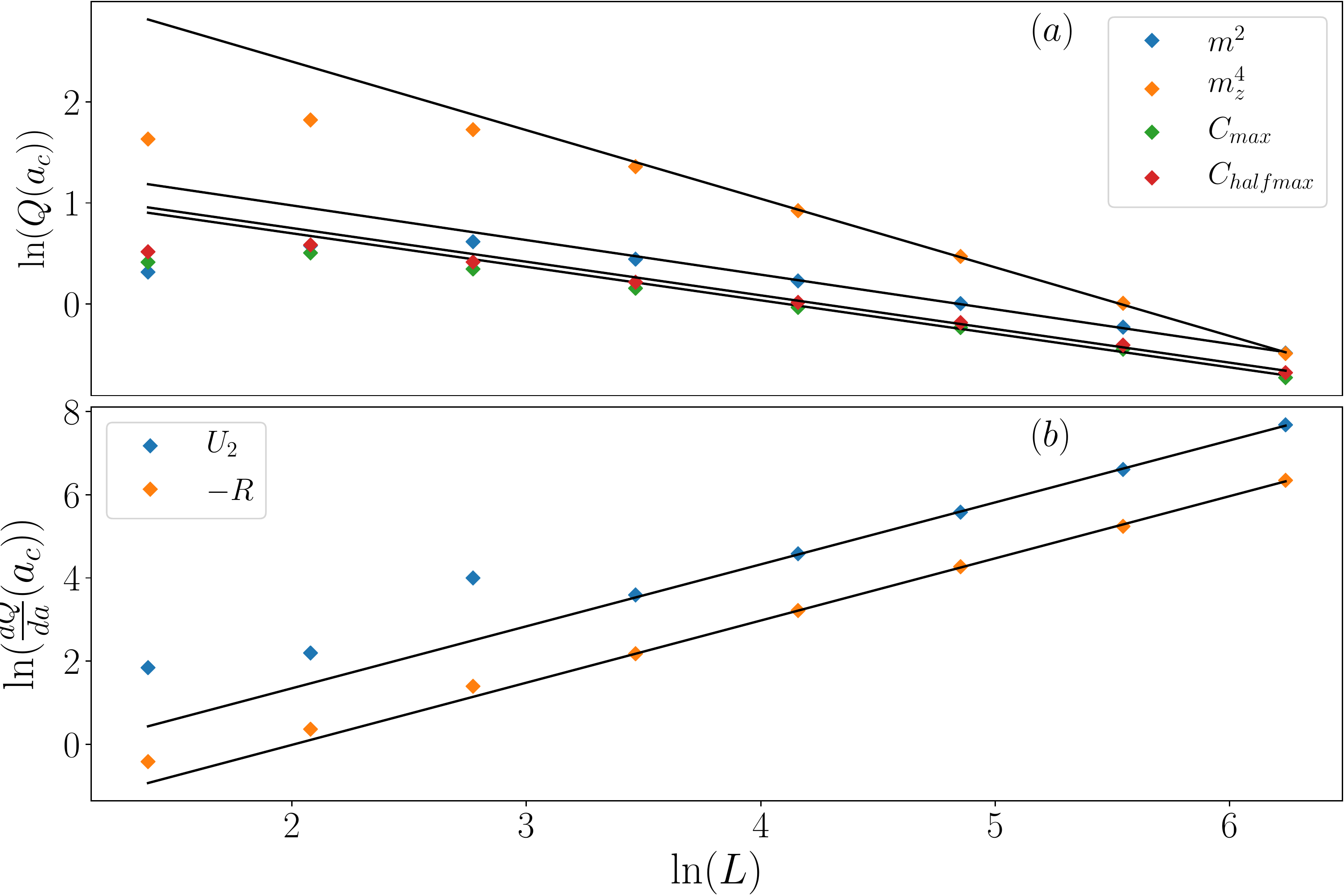}  
  \caption{(Color online) a) $\ln(Q)$ as a function of $\ln(L)$, where $Q=\langle m^2_z\rangle, \langle m^4_z\rangle, C_{\text{max}}, C_{\text{halfmax}}$. 
  Finite-size scaling at critical point. (b) $\ln\left(\frac{dQ}{da}\right)$ as a function of $\ln(L)$, where $Q=U_2$ and $-R$.}
  \label{fig:Q_ac_a0}
\end{figure}

In this appendix we present our results for the AKLT-FM transition at $a_1=0$.
In this case, we vary $a_2$ across the phase boundary and perform the FSS analysis developed in Sec.\ref{AKLT-FM}.
In Fig.~\ref{fig:BSA_a0} (a), (c), (e) we plot $\langle m^2_z\rangle$, $\langle m^4_z \rangle$, and $U_2$ as a function of $a_2$.
By collapsing the data we find $a_c \approx {1.779(6)}$ 
and $1/\nu \approx {1.5(0), 1.5(0), 1.4(9)}$ from $\langle m^2_z\rangle$ ,$\langle m^4_z \rangle$, and $U_2$ respectively.
Furthermore, we find $2\beta/\nu \approx {0.3(2)}$ from $\langle m^2_z\rangle$ 
and $4\beta/\nu \approx {0.6(4)}$ from $\langle m^4_z \rangle$ respectively.
In Fig.~\ref{fig:BSA_a0} (b), (d), (f) we plot the rescaled data and excellent collapse is observed.
In FIg.~\ref{fig:BSA_C_a0} we plot the original and the rescaled $C_{\text{max}}$, $C_{\text{halfmax}}$, and $R$ 
as a function of $a_2$ and $L^{1/\nu} (a_2-a_c)$ respectively. 
Here we find $a_c \approx {1.779(6)}$ and $1/\nu \approx {1.4(7), 1.4(8), 1.4(2)}$ 
from $C_{\text{max}}$, $C_{\text{halfmax}}$, and $R$ respectively.
Furthermore, we find $\eta \approx {0.2(7)}$ from $C_{\text{max}}$ and $C_{\text{halfmax}}$.
From the crossing of $U_2$ and $R$ we also find $a_c \approx {1.779(6)}$.
Finally in Fig.\ref{fig:Q_ac_a0} we plot the log of various quantities at the estimated critical point.
Here we observe a larger deviation from the scaling for smaller sizes data.
By fitting the data for $L \ge 64$ we find 
$2\beta/\nu \approx {0.3(4)}$, $4\beta/\nu \approx {0.6(8)}$, $\eta \approx {0.3(3)}$, $\nu \approx {1.4(9)}$.
We observe that the results are highly consistent with each other. 
However, the estimated critical exponents are clearly not the one for the 2D Ising model.
This confirms that the transition does not belong to the 2D Ising universality class. 


\bibliography{ref}

\end{document}